\documentclass[11pt,a4paper]{article}
\usepackage{amsfonts}
\usepackage[dvips]{graphicx}
\usepackage{amsthm}
\usepackage{amsmath}
\usepackage{epsfig}
\usepackage{t1enc}
\usepackage{amssymb}
\usepackage{latexsym}
\usepackage{bm}
\usepackage[hang]{caption}

\title{\bf From `nothing' to inflation and back again}

\author{Vladim\'ir Balek\footnote{e-mail
address: balek@fmph.uniba.sk}
\\
{\it Department of Theoretical Physics, Comenius University,
Bratislava, Slovakia}}

\begin{document}

\maketitle
\maketitle\abstract

A procedure for solving Wheeler-DeWitt equation in Euclidean
region, following step by step the construction of tunneling wave
function in nonrelativistic quantum mechanics by Banks, Bender and
Wu, is proposed. Solutions for a universe satisfying no-boundary
condition and a universe created from `nothing' are compared to
the corresponding solutions for a particle in a two-dimensional
potential well, and effects of indefiniteness of metric and zero
energy in Wheeler-DeWitt equation are discussed.

\section{Introduction}

The basic two minisuperspace solutions of Wheeler-DeWitt equation
were proposed as long ago as at the beginning of 80's. In 1983,
Hartle and Hawking introduced their no-boundary condition and used
it to construct an inflationary solution in which scalar field was
decoupled from gravity and its effect on the expansion of the
universe was mimicked by the cosmological constant
\cite{hartle-83}. Hawking then replaced the conformal coupling of
the scalar field by the mass term and obtained a truly
inflationary solution \cite{{hawking-84a}, {hawking-84b}}. Since
then the model has been explored repeatedly. To mention just two
remarkable results, the universe was shown to avoid Big Crunch by
quantum bounce for all, not just fine-tuned, initial conditions
\cite{singh-04}; and the model was used to demonstrate the
collapse of the measurement of time with the help of quantum
mechanical degrees of freedom in the period of maximal expansion
\cite{hohn-12}. Shortly before the no-boundary solution appeared,
in 1982, Vilenkin proposed another solution describing creation of
the universe from `nothing' \cite{{vilenkin-82}, {vilenkin-86}}.
Linde advocated the use of such solution `in those situations
where the scale parameter $a$ itself must be quantized'
\cite{linde-90}, and proposed a heuristic derivation of it via
inverse Wick rotation. Later there appeared a different approach
to `creationist' cosmology, adopting the no-boundary condition but
making use of density matrix rather than wave function
\cite{barvinsky-06}.

Both Hawking and Vilenkin solutions contain an Euclidean region in
which the wave function describes tunneling of the universe,
either from a finite radius to a point (Hawking solution) or from
a point to a finite radius (Vilenkin solution). In nonrelativistic
quantum mechanics, a complete WKB solution in the tunneling region
in two dimensions was obtained by Banks, Bender and Wu
\cite{banks-73}. The solution refers to a particle escaping from a
potential well, but with some effort can be modified to apply also
to the opposite case when a particle tunnels from outside the
barrier into the well. The cosmological problem is two-dimensional
and allows for WKB approximation, therefore it would be natural to
employ the same construction in it. However, as for now the
solution by Banks, Bender and Wu was apparently not used in this
way, although Vilenkin mentions in \cite{vilenkin-86} that his
tunneling solution is `similar' to it.

In this note a procedure for computing the wave function of the
universe in the Euclidean region is outlined. In section II the
construction by Banks, Bender and Wu is summarized, rewritten so
that it works both ways, outwards as well as inwards; in section
III new features of the construction appearing when it is carried
over to cosmology are discussed; and in section IV possible
applications of the theory are suggested.

\section{Tunneling in quantum mechanics}

Consider a particle in two dimensions obeying Schr\"odinger
equation
\begin{equation}
  \big[- \partial_x^2 - \partial_y^2 + \mbox{\small $\frac 14$}
  (x^2 + y^2)   - \mbox{\small $\frac 14$} \epsilon (x^4 + y^4 +
  2c x^2 y^2) - E\big] \psi = 0,
  \label{eq:QMeq}
\end{equation}
where the parameters $\epsilon$ and $c$ satisfy $0 < \epsilon \ll
1$ and $|c| < 1$. The potential into which the particle is placed
consists of a circularly symmetric well and a quatrefoil barrier
around it. We are interested in two processes, tunneling of the
particle from the well to the other side of the barrier and {\it
vice versa}, both along the positive $x$ axis. The particle
tunneling outwards has a (quasi)discrete spectrum, which coincides
to a great precision, if the energies are not too high, with the
spectrum in a well extrapolated to infinity. From the requirement
that only an outgoing wave appears behind the barrier it follows
that the energies acquire a small imaginary part equal to $-$
\small{$\frac 12$} $\times$ the decay rate of the state inside the
well. The particle tunneling inwards, on the other hand, can have
any energy. In what follows we assume that $\psi$ separates in
variables $x$, $y$ inside the well and the energy $E_y$ going to
the $y$ direction equals $\frac 12$. In particular, the particle
tunneling outwards can be in the ground state with $E = 1$ and
$E_x = E_y = \frac 12$.

Let us start with rewriting the potential near the $x$ axis as
\begin{equation}
  V = V_0 + \mbox{\small $\frac 14$} k y^2,
  \label{eq:Vx}
\end{equation}
where
\begin{equation}
  V_0 = \mbox{\small $\frac 14$}x^2 (1 - \epsilon x^2), \quad
  k = 1 - 2 \epsilon cx^2.
  \label{eq:V0k}
\end{equation}
The idea of solving (\ref{eq:QMeq}) in the tunneling region is to
separate a WKB wave function in the longitudinal direction and a
bell-shaped function with a variable width in the transversal
direction out of $\psi$. Thus, we write
\begin{equation}
  \psi = A p^{-1/2} \exp \Big(\mp \int pdx - \mbox{\small $\frac 14$}f
  y^2\Big), \quad p = \sqrt{V_0 - E_x},
  \label{eq:pQM}
\end{equation}
with $f$ depending on $x$ and $A$ depending on both $x$ and $y$.
The upper and lower sign refer to a particle tunneling outwards
and inwards respectively. Equation for $f$ is obtained by
collecting the terms in the original equation proportional to
$y^2$ and putting them equal to zero. In this way we relate the
function $f$, defining the width of the wave function, to the
function $k$, defining the width of the valley the particle is
tunneling through; and we obtain an equation for $A$ that
separates after $y$ is replaced by an appropriate variable
proportional to it.

In the leading order of WKB approximation equations for $f$ and
$A$ lose the second derivative with respect to $x$ and, in
addition, the function $p$ multiplying the first derivative loses
the constant $E_x$ under the square root sign. After rescaling $x
\to \epsilon^{-1/2} x$ we have
$$2p \frac d{dx} \to xw \frac d{dx} = - x^2 \frac d{dw},
\quad w = \sqrt{1 - x^2},$$ and equation for $f$ takes the form
\begin{equation}
  \pm x^2 \frac {df}{dw} = f^2 - k, \quad k = 1 - 2cx^2.
  \label{eq:fQM}
\end{equation}
The equation can be linearized by introducing an auxiliary
function $u$ such that
\begin{equation}
  f = \mp \frac {x^2}u \frac {du}{dw}.
  \label{eq:uQM}
\end{equation}
After inserting this into (\ref{eq:fQM}) we find that $u$ obeys
equation for associated Legendre function with the lower index
given by
\begin{equation}
  \nu (\nu + 1) = 2c
  \label{eq:n}
\end{equation}
and the upper index equal to $\pm 1$. Equation for $A$ separates
in the variables $w$ and $s = y/u$ and if we write $A$ as $WS$,
where $W$ is a function of $w$ and $S$ is a function of $s$, we
find that $S$ is either cosine or hyperbolic cosine or constant.
Matching the tunneling solution to the solution inside the well
singles out the third possibility. Thus, $A$ depends on $w$ only,
and after a little algebra one finds
\begin{equation}
  A = const\ \Big(\frac {1 - w}{1 + w}\Big)^{\pm 1/4} u^{-1/2}.
  \label{eq:A}
\end{equation}

Let us examine the behavior of $f$. To simplify the analysis, we
return from the variable $w$ to $x$ with the help of the formula
$$x \frac d{dw} = - w \frac d{dx}.$$
First we find the asymptotic of $f$ for $x \sim 0$. In order that
the internal and external solutions match, $f$ must equal 1 at $x
= 0$; thus, we write $f$ as $1 + \Delta$ and skip the $\Delta^2$
term in (\ref{eq:fQM}). We obtain the equation
\begin{equation}
  \mp x \frac {d\Delta}{dx} = 2(\Delta + cx^2),
  \label{eq:D}
\end{equation}
whose solution is
\begin{equation}
  \Delta = \bigg \{ \mbox{\hskip -2mm \large
  $\left. \begin{array} {l}
  \mbox{\small $-\frac 12 cx^2$}\\
  \mbox{\small $2c x^2(\log x + {\cal C})$}\\
  \end{array}\mbox{\hskip -1mm}, \right.$}
  \label{eq:Dsol}
\end{equation}
where the upper and lower expression refer to the upper and lower
sign in (\ref{eq:D}) and $\cal C$ is integration constant. We can
see that the width of the tunneling wave function is given
uniquely for a particle tunneling outwards, but the strips on
which the wave function is nonzero form a one-parametric sequence
for a particle tunneling inwards. The reason is presumably that
the waves incident on the barrier along the positive $x$ axis form
a one-parametric sequence, too, and different waves tunnel along
different stripes.

From the asymptotic of $f$ we can determine $u$. For $w \sim 1$,
or equivalently, $x \doteq \sqrt{2(1 - w)} \sim 0$, the
asymptotics of the associated Legendre functions of first and
second kind are
$$P_\nu^{-1} = \mbox{\small $\frac 12$} x, \quad Q_\nu^1 = -
x^{-1}.$$ (It suffices to consider one upper index in $P_\nu$ and
$Q_\nu$ since the functions with opposite upper indices are
proportional to each other.) To obtain $f = 1$ for $x = 0$ we must
choose $u$ equal to $P_\nu^{-1}$ and to $Q_\nu^1$ $+$ some
coefficient $\times$ $P_\nu^{-1}$ for a particle tunneling out-
and inwards. Moreover, by including the next-to-leading term into
the expansion of $Q_\nu^1$,
$$\Delta Q_\nu^1 = \mbox{\small $\frac 14$} \nu (\nu + 1) \big[2
\log \mbox{\small $\frac x2$} + \psi(\nu) + \psi(\nu + 2) -
\psi(2) + \gamma \big] x + \mbox{\small $\frac 14$} x,$$ where
$\psi$ is digamma function and $\gamma$ is Euler-Mascheroni
constant, we can find the coefficient in front of $P_\nu^{-1}$ in
the latter case.

Finally, we can use $u$ to compute $f$ on the whole interval of
$x$. Note that for $c > \frac 12$ the function $k$ becomes
negative before $x$ reaches 1, so that the valley the particle is
tunneling through turns to a slope. According to (\ref{eq:fQM}),
this drives the derivative $df/dx$ to more negative values and
makes $f$ to fall down faster for a particle tunneling outwards.
Nevertheless, $f$ stays positive and $\psi$ stays suppressed in
the transversal direction up to $x = 1$ for any $c$. If the
particle is tunneling inwards, $f$ increases near $x = 1$ for the
values of $c$ in question, and only with decreasing $c$ it starts
to decrease. For low enough $c$'s $f$ passes through zero before
$x$ reaches 1; however, this holds only for ${\cal C} = 0$ and can
be cured by choosing positive $\cal C$.

\section{Tunneling in cosmology}

Consider a closed universe with a scalar field living in it, and
denote the radius of the universe by $a$ and the value of the
field by $\phi$. Suppose, furthermore, that the field is massive
with the mass $m$ and the theory includes cosmological constant
$\Lambda$; thus, the field has quadratic potential shifted upwards
by $\lambda =$ \small{$\frac 13$}$\Lambda$. The wave function of
the universe satisfies the Wheeler-DeWitt equation
\begin{equation}
  \big[- \partial_a^2 + a^{-2}\partial_\phi^2 + a^2 (1 - \lambda a^2 -
  m^2 a^2 \phi^2)\big] \psi = 0.
  \label{eq:ceq}
\end{equation}
To account for the ambiguity due to operator ordering, the
operator $-\partial_a^2$ has to be modified to $- a^{-\mu}
\partial_a a^\mu \partial_a$ with an arbitrary $\mu$. However,
in the WKB approximation we are interested in this affects only
the preexponential factor in $\psi$, therefore we can put $\mu =
0$.

Equation (\ref{eq:ceq}) is almost identical to the equation we
obtain from (\ref{eq:QMeq}) by passing from rectangular to polar
coordinates, restricting the polar angle to $|\varphi| \ll 1$ and
replacing the operator $- r^{-1} \partial_r r \partial_r$, again
with the reference to WKB approximation, by $-\partial^2_r$. The
new equation reads
\begin{equation}
  \big[- \partial_r^2 - r^{-2} \partial_\phi^2 + \mbox{\small
  $\frac 14$} r^2 (1 - \epsilon r^2 + 2\epsilon \gamma r^2 \varphi^2)
  - E\big] \psi = 0,
  \label{eq:QMeqr}
\end{equation}
where $\gamma = 1 - c$. We will solve this equation in a similar
way as equation (\ref{eq:QMeq}) and discuss equation
(\ref{eq:ceq}) later. First we express $\psi$ as
\begin{equation}
\psi = B q^{-1/2} \exp \Big(\mp \int q dr - \mbox{\small $\frac
14$}g \varphi^2\Big), \quad q = \sqrt{{\cal V}_0 - E},
  \label{eq:pQMr}
\end{equation}
with ${\cal V}_0 = \frac 14 r^2 (1 - \epsilon r^2)$ and $g$
related to $\kappa = 2\epsilon \gamma r^4$ in a similar way as $f$
was related to $k$. After rescaling $r \to \epsilon^{-1/2} r$,
$\varphi \to \epsilon^{1/2} \varphi$, $\kappa \to \epsilon^{-1}
\kappa$ and $g \to \epsilon^{-1} g$ and introducing $\xi = \sqrt{1
- r^2}$ we have
\begin{equation}
\pm r^2 \frac {dg}{d\xi} = r^{-2} g^2 - \kappa, \quad \kappa =
2\gamma r^4,
  \label{eq:gQM}
\end{equation}
and after writing $g$ as
\begin{equation}
g = \mp \frac {r^4}v \frac {dv}{d\xi},
  \label{eq:vQM}
\end{equation}
we find that $v$ obeys equation for Gegenbauer function with the
lower index given by
\begin{equation}
  \alpha (\alpha + 3) = - 2\gamma
  \label{eq:a}
\end{equation}
and the upper index equal to $\frac 32$. Finally, for the function
$B$ we obtain
\begin{equation}
B = const\ v^{-1/2}.
  \label{eq:B}
\end{equation}

To fix the combination of Gegenbauer functions in $v$ we must
explore the behavior of $g$ for $r \sim 0$. When doing so we
notice that now there are {\it two} terms proportional to $y^2$ in
the exponential near the origin, the term $-\frac 14 g \varphi^2
\doteq -\frac 14 r^{-2} g y^2$ and the term $\mp \frac 14 y^2$
coming from
$$\int q d\hat r \sim \int \mbox{\small $\frac
12$} \hat r d\hat r = \mbox{\small $\frac 14$} \hat r^2 =
\mbox{\small $\frac 14$} (\hat x^2 + y^2),$$ where we have denoted
the original, non-rescaled variables $x$ and $r$ by $\hat x$ and
$\hat r$. These two terms must add to produce the term $-\frac 14
y^2$ appearing in the solution inside the well. The resulting
asymptotics of $g$ are
\begin{equation}
  g = \bigg \{ \mbox{\hskip -2mm \large
  $\left. \begin{array} {l}
  \mbox{\small $\frac 12 \gamma r^4$}\\
  \mbox{\small $2r^2[1 + cr^2(\log r + {\cal D})]$}\\
  \end{array}\mbox{\hskip -1mm}. \right.$}
  \label{eq:gsol}
\end{equation}
This yields $v$ equal to $C_\alpha^{3/2}$ and $D_\alpha^{3/2}$ $+$
some coefficient $\times$ $C_\alpha^{3/2}$ for a particle
tunneling out- and inwards respectively; and knowing $v$, we can
determine the behavior of $g$ on the whole interval of $r$.

Once we have found $f$ we do not need to compute $g$ from scratch.
Instead, we can express $g$ in terms of $f$. For that purpose we
insert
$$\int q d\hat r = \int \mbox{\small $\frac
12$} \hat r \xi d\hat r = \epsilon^{-1} \int \mbox{\small $\frac
12$} r \xi d r = - \epsilon^{-1} \mbox{\small $\frac 16$} (\xi^3 -
1) \doteq \int p d\hat x + \mbox{\small $\frac 14$} \xi y^2$$ into
(\ref{eq:pQMr}) and compare the resulting expression to
(\ref{eq:pQM}). We obtain
\begin{equation}
g = r^2 (f \mp \xi),
  \label{eq:gf}
\end{equation}
where $f$ is to be regarded as a function of $r$. This coincides
with the function $g$ constructed previously, if we express
$C_\alpha^{3/2}$ and $D_\alpha^{3/2}$ in terms of $P_\nu^{-1}$ and
$Q_\nu^1$ and put ${\cal D} = {\cal C} - \frac 14$.

Equation (\ref{eq:ceq}) differs from (\ref{eq:QMeqr}) in that that
it has reversed sign of the kinetic term in $\phi$ direction and
vanishing energy. Because of the former property the metric of the
kinetic term is indefinite. The sign of the mass term is reversed
or stays the same depending on whether $m$ is real, which
corresponds to a metastable vacuum, or imaginary, which
corresponds to an unstable vacuum. However, it is the {\it
relative} signs with respect to the kinetic terms in $a$ and
$\phi$ directions which matter; and no matter what the absolute
sign, one of these signs stays the same while the other is
reversed.

The two solutions of equation (\ref{eq:ceq}), with minus and plus
sign in the exponential, describe tunneling of the universe
outwards, from a point to a finite radius, and inwards, from a
finite radius to a point. An immediate consequence of the
indefinitness of metric is that the signs in the equation for $g$
are switched. As a result, we obtain one-parametric class of $g$'s
for a universe tunneling outwards and a single $g$ for a universe
tunneling inwards. Having just one solution in the latter case is
consistent with the observation that the wave function of the
universe is determined completely by the no-boundary condition.
Having infinitely many solutions in the former case can be
explained by the fact that the energy is zero, which means that
the imaginary part of energy is zero, which means that the
outgoing probability current stays finite up to the zero radius of
the universe. The corresponding outgoing waves must be put into
the theory by hand and apparently form a one-parametric sequence,
similarly as ingoing waves in the quantum mechanical problem with
a particle tunneling inwards.

In the previous discussion we have assumed that $m$ is real.
However, as mentioned by Vilenkin in \cite{vilenkin-86}, when
considering a universe tunneling outwards it is reasonable to pass
to imaginary $m$. The point is that the tunneling path shortens,
and the wave function becomes less suppressed behind the barrier,
if the shift of the potential $\lambda$ increases. Thus, the
tunneling is most effective at the global maximum of the
potential. Imaginary $m$ and tunneling outwards in cosmology
corresponds to $c > 1$ and tunneling inwards in quantum mechanics.
For such tunneling we find, in addition to a one-parametric class
of $g$'s with the asymptotic given in the lower line of
(\ref{eq:gsol}), one more $g$ with the asymptotic
\begin{equation}
g = - \mbox{\small $\frac 12$} \gamma r^4.
  \label{eq:gsol1}
\end{equation}
This solution must be abandoned since the corresponding $f$ equals
$-1$ at $x = 0$, which means that the wave function explodes in
the transversal direction. However, in the cosmological setting
such $g$ seems admissible, and even privileged because of its
one-to-one correspondence with $g$ appearing in the problem with a
universe tunneling inwards.

\section{Conclusion}

We have shown how the procedure by Banks, Bender and Wu can be
carried over to cosmology and used to construct the wave function
of the universe in the Euclidean region. The way how to do that
has been only sketched here, the details will be given elsewhere
\cite{balek-12}. In particular, we have skipped the discussion of
the behavior of the wave function `inside the well', in the region
where the radius of the universe is close to zero. This question
is vital for the construction, because as long as the tunneling
solution is not matched with the solution `inside the well', it
remains unjustified.

For a universe tunneling inwards, tunneling solution converts at
the edges of Euclidean region into oscillatory one, describing
time-symmetric evolution during which the universe repeatedly
crosses that region. The crossings do not seem to change the
course of the evolution, but can still have some imprint on it,
and the knowledge of the exact form of tunneling solution can help
to identify that imprint. For a universe tunneling outwards there
apparently exists, in addition to `regular' solution which behaves
like that for a universe tunneling inwards, a one-parametric class
of solutions with markedly different behavior. Such solutions can
mediate tunneling not only to the maximum of potential but also to
its minimum, therefore can be helpful when contemplating the
possibility of creating a universe with scalar field in metastable
state directly from 'nothing'.

\end{document}